\documentclass[lettersize,journal]{IEEEtran}
\usepackage{amsmath,amsfonts}
\usepackage{algorithmic, upgreek, float, bm, cite}
\usepackage{array}
\usepackage[caption=false,font=normalsize,labelfont=sf,textfont=sf]{subfig}
\usepackage{textcomp}
\usepackage{stfloats, amssymb,xcolor}
\usepackage{url}
\usepackage{verbatim}
\usepackage{graphicx}
\def\BibTeX{{\rm B\kern-.05em{\sc i\kern-.025em b}\kern-.08em
    T\kern-.1667em\lower.7ex\hbox{E}\kern-.125emX}}

\newcommand{\vect}[1]{\mathbf{#1}}
\newcommand{\mathvec}[1]{\boldsymbol{#1}}
\newcommand{\vectrho}{\bm{\rho}}
\newcommand{\appenref}[1]{Appendix A}
\newcommand{\secref}[1]{Sec.~\ref{#1}}

\newcommand{\figref}[1]{Fig.~\ref{#1}}

\newcommand{\myedit}[1]{{#1}}
\newcommand{\myremove}[1]{}

\usepackage{balance}
\begin{document}
\title{Inverse-Designed Tapers for Compact Conversion Between Single-Mode and Wide Waveguides}
\author{Michael~J.~Probst, Arjun~Khurana, Archana~Kaushalram, and Stephen~E.~Ralph
\thanks{The authors are with Georgia Institute of Technology, Atlanta,
GA 30332, USA. This research was supported in part through research cyberinfrastructure resources and services provided by the Partnership for an Advanced Computing Environment (PACE).  This material is based upon work supported in part by the National Science Foundation (NSF) Center ``EPICA'' under Grant No. 2052808, https://epica.research.gatech.edu/.}}

\maketitle

\begin{abstract}
Waveguide tapers are critical components for leveraging the benefits of both single-mode and wide waveguides. Adiabatic tapers are typically hundreds of microns in length, dramatically limiting density and scalability. We reenvision the taper design process in an inverse-design paradigm, introducing the novel L-taper. We present a novel approach to inverse-designed tapers where the input and output waveguides are rotated $\mathvec{90^\circ}$ with respect to each other. The resultant design has an order-of-magnitude smaller footprint, and the design process is compatible with a variety of fabrication processes. We demonstrate an L-taper designed on 220~nm silicon-on-insulator that converts a 0.5~$\mathvec{{\upmu}}$m waveguide to a 12~$\mathvec{\upmu}$m waveguide with $\mathvec{-0.38}$~dB transmission and 40~nm 1-dB bandwidth. The footprint is $\mathvec{16\ \upmu}$m $\mathvec{\times6\ \upmu}$m, representing a 12\texttimes~smaller footprint than a linear taper with the same transmission.
\end{abstract}


\section{Introduction}

\IEEEPARstart{H}{igh} aspect ratio photonic waveguides are critical components in a variety of integrated photonic applications, including high power, low loss routing \cite{john2012multilayer,heck2014ultra,Xiang:22}; multimode photonics \cite{yang2022multi,xu2020silicon}; and out-of-plane transmission (e.g. grating couplers) \cite{taillaert2004compact,cheng2020grating}. On the other hand, single-mode integrated photonics provides other inherent benefits, including high device density and zero modal dispersion. Therefore, leveraging the benefits of both the single-mode waveguides and high aspect ratio waveguides in a unified platform will yield high-performance, high-density integrated photonic systems desirable for a variety of communications, sensing, and quantum applications \cite{li2019multimode}. Such a unified platform requires tapers that interconnect the single-mode and high aspect ratio waveguides. To this end, there exist several design methodologies, each with its own set of tradeoffs. Perhaps the simplest design is the adiabatic taper, in which a linear, logarithmic, or parabolic profile is used to gradually expand the fundamental waveguide mode into a larger guide. This method enjoys ultra-wide bandwidths, straightforward design, and high conversion efficiencies \cite{liang2022adiabatic}. However, adiabatic tapers also dramatically limit device packing density and scalability, as these structures are hundreds of microns or even millimeters in length \cite{cheng2020grating}. Also, in platforms with non-negligible loss, such as thin-film lithium niobate, adiabatic tapers suffer from excess losses\cite{tfln_loss}, curtailing some of the benefits of gradual modal expansion\myedit{/contraction}. Other designs leverage lens-based optics to perform rapid mode \myedit{conversion} \cite{badri2019ultrashort,Abbaslou:17}. This method, while compact and efficient, often requires custom material platforms and less-than-scalable fabrication techniques. A more effective approach is to utilize inverse-design algorithms which can perform rapid \myedit{mode conversion} on conventional material platforms with standard, high-volume nanofabrication techniques by producing non-intuitive geometries \cite{probst2023topology,garza2022fast,Michaels:18}. Inverse-designed tapers are compact, typically $\mathbin{\sim}{3}$\texttimes~the width of the wider waveguide in length. Furthermore, mature algorithms leverage several layers of parallelism to manage the inherent computational complexity of the design process, while robust optimization produces fabrication-robust devices \cite{hammond2022high,hammond_constraints,zhou_geometric}.

In this work, we present a novel approach to inverse-designed tapers where the input and output waveguides are rotated $90^\circ$ with respect to each other which dramatically decreases the device footprint \myedit{beyond what was formerly achievable by inverse-design methodologies}. The design problem now resembles a vertical grating coupler; however, unlike the grating coupler that heavily relies on an asymmetric stackup (e.g. partial-/multi-etch or multiple layers) to direct the wavevector \myedit{out-of-plane}, the optimizer utilizes thousands or millions of voxels to direct the light \myedit{from the single-mode waveguide toward} the \textit{in-plane} wide waveguide \myedit{(and by reciprocity from the wide waveguide toward the single-mode waveguide)}. The resultant geometry, herein called an L-taper, is shown to produce near-unity transmission with ultra-compact footprint (\figref{fig:performance}b). The L-taper can convert from a single-mode waveguide to arbitrarily wide waveguides (e.g., tens of microns) with a length of approximately the width of the waveguide, whereas linear tapers require increasingly large footprints for expansion to such wide waveguides. The device, being a Bragg-like structure, has bandwidth of tens of nanometers, which is more than sufficient for conventional telecom applications \cite{lal2016full}. Finally, the L-taper design technique admits a variety of material platforms, geometric lengthscale constraints, and robust optimization techniques which are prevalent throughout inverse-design \cite{hammond_constraints,probst2024fabrication,zhou_geometric}.

\begin{figure}
    \centering
    \includegraphics[width=3.49in]{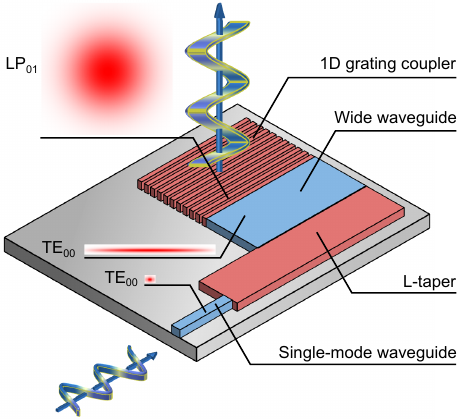}
    \caption{A single-mode waveguide is rapidly expanded to a wide waveguide using the novel L-taper and then connected to a fiber grating coupler. All components can be inverse-designed in a compact footprint.}
    \label{fig:motivation}
\end{figure}

In \secref{sec:opt}, we describe the design via topology optimization (TO) of an L-taper to convert a $0.5\ \upmu$m waveguide to a 12 $\upmu$m waveguide with $-0.38$~dB transmission on a 220~nm silicon-on-insulator (SOI) process. 
12 $\upmu$m waveguides are crucial for integrated photonics because the mode profile has high overlap with that of a standard single-mode fiber (SSMF). \myedit{Thus, a grating coupler design strategy is to design two devices: an in-plane taper that converts a single-mode waveguide to a wide waveguide and a 1D grating coupler that couples from the wide waveguide to an optical fiber (\figref{fig:motivation}). Separating the beam expansion into distinct functional areas permits overall higher coupling and simplifies the functionality of each device \cite{cheng2020grating}.} In \secref{sec:discussion}, we compare the novel L-taper with linear tapers, substantiating that the L-taper has a 12\texttimes~footprint reduction for similar coupling efficiency \myedit{and examining the higher order modal excitation in the waveguide for both a $100\ \upmu$m taper and the novel L-taper.} \myremove{We demonstrate the L-taper has low excitation of higher order modes in the wide waveguide and also examine fabrication tolerance by performing an over-/under-etch analysis on the transmission of the taper.}

\section{Device optimization} \label{sec:opt}
The L-taper was designed using density-based TO \cite{sigmund_overview,jensen2011topology,christiansen2021inverse}, a subset of inverse-design where the design region is parameterized by individual pixels which continuously evolve between the design materials using gradient-based optimization algorithms. The design pixels are gradually forced to binary states such that the design can be fabricated. We formulate device design with the optimization problem:
\begin{equation}\label{eq:obj_old}
    \begin{matrix}
         & \min\limits_{\vectrho}\big\{f_n(\vect{E})\big\} & n\in\left\{1,2,3\right\} \\
         \text{s.t.} & \nabla\times\frac{1}{\mu}\nabla\times\vect{E}-\omega_m^2\vect{\varepsilon}(\vectrho)\vect{E}=\ -j\omega_m\vect{J} & m\in\left\{1,2,3\right\} \\
         &0\leq\vectrho\leq1 & \\
         & g_k(\vectrho)\leq0 & k\in\{1,2\}
    \end{matrix}
\end{equation}
where $\vect{E}$ is the electric field, $\vect{J}$ is the source current, and $\vectrho$ is the design variables which determine the permittivity matrix $\varepsilon$ using a filter-threshold projection function \cite{sigmund_overview}. The figure of merit (FOM), $f_n$, is defined at each wavelength as:
\begin{equation}
    f_n(\vect{E})=10\times\log_{10}\left(\left|\cfrac{\alpha_{1}}{\alpha_0}\right|^2\right)
\end{equation}
where $\alpha_0$ and $\alpha_1$ correspond to the forward propagating mode overlaps of the fundamental mode of the input and output waveguides, but an arbitrary phase profile in the wide waveguide could also be achieved. The device was optimized at $\lambda=$ 1540~nm, 1550~nm, and 1560~nm which ensured high transmission across C-band. Finally, the geometry is subject to linewidth and linespacing constraints, $g_1$ and $g_2$, which enforce a minimum feature size \cite{zhou_geometric}. For a comprehensive description of the parameterization techniques and design process, the readers are directed to \cite{probst2024fabrication}. Throughout this work, the open-source finite-difference time domain solver Meep \cite{meep} was used to evaluate device performance and the built-in hybrid time/frequency adjoint variable method solver \cite{hammond2022high} calculated the gradient of the FOM with respect to all design parameters at all design wavelengths.

\begin{figure*}[htb]
    \centering\includegraphics[width=7in]{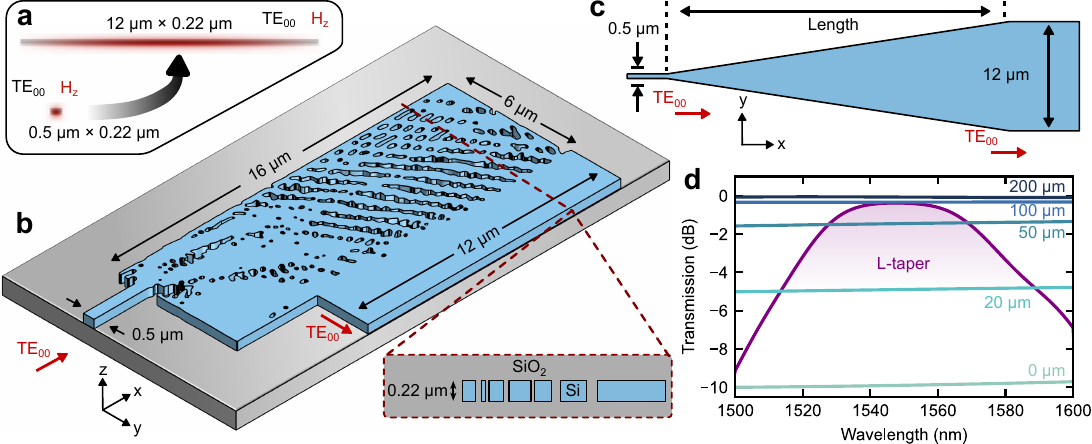}
    \caption{(a) The topology optimized L-taper rapidly converts between the fundamental mode of a 0.5~$\upmu$m waveguide and the fundamental mode of a 12~$\upmu$m waveguide on a 220~nm SOI process. (b) The design problem is initialized with the input and output waveguides rotated $90^\circ$ relative to each other and a $16\ \upmu\text{m}\times6\upmu\text{m}$ design region. The optimized structure employs apodized slanted gratings to rapidly expand/contract the mode profile with high efficiency. (c) In contrast, linear tapers are designed by gradually increasing the width of the waveguide, which requires 12\texttimes~larger footprint and dramatically limits the scalability. (d) The inverse-designed taper performs similar to a $100\ \upmu\text{m}$ taper at peak coupling efficiency, though the linear tapers demonstrate significantly larger bandwidth.}
    \label{fig:performance}
\end{figure*}

We designed a taper using a 220~nm silicon-on-insulator (SOI) process clad with SiO$_2$ and a 50~nm feature size to convert the fundamental (TE$_{00}$) mode of a 0.5 $\upmu$m waveguide to the fundamental mode of 12~$\upmu$m waveguide (\figref{fig:performance}a). The design was simulated at a resolution of $30\ \text{voxels}/\upmu$m, and the optimization was performed with a resolution of $60\ \text{voxels}/\upmu$m. The design region was $16\ \upmu$m $\times\ 6\ \upmu$m which corresponds to  $\mathbin{\sim}{350,000}$ design parameters. Extending the design region 4~$\upmu$m beyond the edge of the large waveguide (\figref{fig:performance}b) led to an overall increase in coupling efficiency \myedit{because the wavefront from the single-mode waveguide expanded before reaching the slanted gratings}. Since the stackup is vertically symmetric about the design layer, we inserted a symmetry plane in the middle of the 220~nm Si layer which halved the computation complexity of the simulation. The taper was optimized over 490 iterations and achieved $-0.38$~dB maximum coupling efficiency at $\lambda=1547$~nm and 40~nm 1-dB bandwidth.

\section{Discussion}\label{sec:discussion}
In this section, we analyze the performance of the L-taper \myedit{in comparison to linear tapers}. First, we compare the transmission of the L-taper to linear tapers ranging from lengths of 0~$\upmu$m to 200~$\upmu$m (\secref{subsec:linear}). Then, we examine the excitation of the higher order TE modes in the wide waveguide (\secref{subsec:multimode}) \myedit{from the L-taper and linear tapers  which are injected with the fundamental TE mode from the single-mode waveguide}. \myremove{Finally, we perform an over-/under-etch analysis of the taper, simulating the design parameters after an erosion and dilation (\secref{subsec:variability}).}

\subsection{Transmission Analysis}\label{subsec:linear}
To contextualize the compact footprint of the L-taper, we evaluate the length needed for a linear taper to achieve the same transmission as the inverse-designed L-taper. Linear tapers \myedit{gradually increase the core width} to connect the small waveguide and the large waveguide (\figref{fig:performance}c). Sufficiently long linear tapers adiabatically expand the mode; however, for short linear tapers, some power is radiated or coupled into higher order modes of the wide waveguide which reduces overall transmission. \figref{fig:performance}d shows the coupling efficiency of linear tapers of various lengths. For a 0 $\upmu$m taper (i.e. butt-coupled), the transmission is $-10$~dB, while the transmission is $-0.2$~dB for a 200~$\upmu$m taper. The L-taper, with peak transmission $-0.38$~dB corresponds to a linear taper of length 100~$\upmu$m (\figref{fig:performance}d). The footprint of a $100\ \upmu\text{m}$ linear taper is $12\ \upmu\text{m}\times100\ \upmu\text{m}=1200\ \upmu\text{m}^2$ in comparison to the L-taper with a footprint of $16\ \upmu\text{m}\times6\ \upmu\text{m}=96\ \upmu\text{m}^2$, a 12\texttimes~reduction in footprint. Such compact footprints of inverse-designed L-taper enable high packing density and mitigate additional losses.  

\subsection{Modal Analysis}\label{subsec:multimode}
\begin{figure}[htb]
    \centering
    \includegraphics{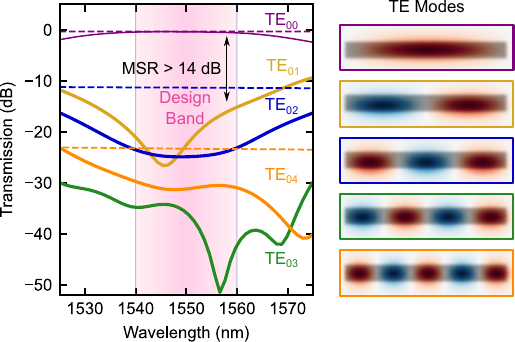}
    \caption{The L-taper (solid lines) and linear taper (dashed lines) transmission to the first five TE modes. Across the design band (1540~nm-1560~nm), there is greater than 14~dB MSR for the L-taper. The MSR and peak transmission can be improved by decreasing the width of the design band and specifically maximizing the modal extinction during the optimization. \myedit{The linear taper has flat 12~dB MSR across the design band. Note that the odd-symmetry modes are not excited in the linear taper due to the symmetry of the geometry.}}
    \label{fig:modes}
\end{figure}

\myedit{Higher order mode excitation in wide waveguides is detrimental to the performance of integrated photonic systems because it reduces power in the desired mode and introduces modal dispersion that degrades transmitted signals. Therefore, tapers must be designed to minimize excitations in the higher order modes. Adiabatic tapers overcome this challenge by performing waveguide expansion slow enough that the optical power remains confined to the fundamental mode, resulting in low loss and no modal dispersion. However, there is no defined mode structure in the inverse-designed L-taper, and we therefore characterize the higher order mode excitation.} We simulated the coupling to first five TE modes (TE$_{00}-$  TE$_{04}$) under single-mode injection to the L-taper. The solid lines in \figref{fig:modes} demonstrate that the mode suppression ratio (MSR) is greater than 14~dB in the design band ($1540\text{~nm}-1560\text{~nm}$), \myedit{with the best MSR being $\mathbin{\sim}{25}$~dB at 1545~nm.} The grating demonstrated higher modal extinction near the center of the design band, so taper could achieve higher peak transmission and MSR with a more narrowband optimization. \myedit{The dashed lines in \figref{fig:modes} represent the higher order modal excitation of the $100\ \upmu$m linear taper. The linear taper achieves MSR of 12~dB, and the higher order modal excitation is exceedingly broadband.}

\section{Conclusion}
The novel inverse-designed L-taper produces equivalent coupling to a linear taper in a 12\texttimes~smaller footprint. This is accomplished by rotating the input and output waveguides $90^\circ$, allowing the design to function like an in-plane grating coupler. \myedit{The design technique is versatile, compatible with a variety of material platforms and with inverse-design fundamentals such as minimum feature size enforcement, multi-frequency objective functions, and gradient-based optimization techniques.} We demonstrated the L-taper by converting the fundamental mode of a $0.5\ \upmu$m waveguide to the fundamental mode of a $12\ \upmu$m waveguide with a $16\ \upmu\text{m}\times6\ \upmu\text{m}$ footprint. The optimized design had $-0.38$~dB transmission and 40~nm 1-dB bandwidth centered at 1547~nm.

\myedit{Future work includes improving the device performance and extending this technique to additional design problems. Reducing the rotation of the two waveguides to less than $90^\circ$ (akin to vertical grating couplers that emit at an angle) may improve performance and hyperparameter optimization of the design region geometry may enable the optimizer to find a more optimal local minimum. Further designs include L-tapers designed to excite a particular higher order mode or arbitrary phase profile in the wide waveguide \myedit{and designs that leverage symmetries to supress higher mode excitation}.} 

\bibliography{references}

\end{document}